\documentclass[aps,prl,twocolumn,showpacs,superscriptaddress]{revtex4-1}
\usepackage{latexsym}
\usepackage{amssymb}
\usepackage{graphicx}
\usepackage{amsmath}
\usepackage{bm}
\usepackage[colorlinks,
          linkcolor=black,
            citecolor=black,
            urlcolor=blue
           ]{hyperref}
\usepackage{verbatim}
\usepackage{mathrsfs}
\usepackage{extarrows}
\usepackage{comment}
\usepackage{mathtools,slashed}

\allowdisplaybreaks[4]

\usepackage{cancel}

\usepackage{amsthm}

\usepackage[OT1]{fontenc}

\begin{document}

\title{Gaplessness indicator by topologically trivial twisting operators}

\author{Yuan Yao}
\email{Corresponding author: smartyao@sjtu.edu.cn}
\thanks{These authors contributed equally to this work.}
 \affiliation{Institute of Condensed Matter Physics, School of Physics and Astronomy, Shanghai Jiao Tong University, Shanghai 200240, China}

\author{Linhao Li}
\thanks{These authors contributed equally to this work.}
\affiliation{Department of Physics, the Pennsylvania State University, University Park, Pennsylvania 16802, USA}

\author{Feng-Feng Song}
\affiliation{Institute for Solid State Physics, The University of Tokyo, Kashiwa, Chiba 277-8581, Japan}

\begin{abstract}
We propose several general necessary conditions for quantum many-body system in one dimension respecting U(1) symmetry to be gapped.
We show that the ground-state expectation value of topologically trivial twisting operators must approach unity in the thermodynamic limit with a certain finite-size scaling.
Equivalently,
its violation can indicate gaplessness of U(1)-symmetric Hamiltonians.
The topological triviality of such a twisting operator enables us to derive infinitely many other gaplessness indicators by static structure factor to any order in real experiments,
which are impossible to obtain by earlier topologically nontrivial twisting operators.
We also apply analytic and numerical calculations to test the efficiency and consistency of our results.

\end{abstract}


\maketitle

\paragraph{Introductions.---}
Identification and characterization of various quantum phases is one of the most essential but difficult tasks in condensed-matter and statistical physics~\cite{Landau:1937aa}.
Based on the low-energy spectrum,
many-body systems can be classified into gapped and gapless phases as a basic step.
A system is called being gapped,
if there is a nonzero energy gap above its finitely degenerate ground state(s) \textit{in the thermodynamic limit;}
otherwise,
the system is gapless~\cite{Zeng:2015aa,zeng2019quantum}.
Thus,
the notion of spectral gap is genuinely a many-body concept which makes the numerical judgement technically difficult due to complicated interactions and large numbers of degrees of freedom.

Quite often,
symmetry has played an essential role in the gaplessness recognization.
The famous Lieb-Schultz-Mattis (LSM) theorem~\cite{Lieb:1961aa} and its extensions~\cite{OYA1997,Oshikawa:2000aa,Hastings:2004ab,NachtergaeleSims} are one of the most notable concepts,
which state that the system satisfying some microscopic data,
e.g., filling and spin-type,
cannot have a gapped and unique ground state once certain symmetry is respected by the Hamiltonian.
It is a model-independent constraint based only on symmetry data,
but it cannot further distinguish gapplessness and being gapped with a nontrivial ground-state degeneracy or give any nontrivial prediction if the system does not fulfill the required microscopic condition.
Recently,
several gaplessness criteria by \textit{topologically nontrivial} twisting operators~\cite{Lieb:1961aa,OYA1997,Resta:1998aa,Resta:1999aa,Aligia:1999aa,Oshikawa:2000aa,Nakamura:2002aa,Hetenyi:2019aa,Hetenyi:2020aa,Aligia:2023aa,Su:2024aa,Tasaki:2018aa} through their expectation values~\cite{Su:2025aa,Tada:2026aa,Cheng:2026kch} have been proposed.
However,
some of these results strongly rely on those strong symmetry structure,
e.g., SU(2) symmetry,
and are inapplicable to the more common U(1)-symmetric systems in condensed matter physics.
The development
applicable to U(1)-symmetric system
needs the prerequisite knowledge of charge filling.
Therefore,
efficient and general criteria by operator expectation values,
to diagnose gaplessness,
to be called ``gaplessness indicators'',
of systems with weaker symmetry,
e.g., U(1) symmetry,
but without further requirements still remain as an open question.

In this Letter,
we propose a gaplessness indicator by the ground-state expectation value of \textit{topologically trivial} twisting operator ${U}_F$ where $F$ can be any winding-free function satisfying certain general conditions.
We show that if a U(1)-symmetric chain with length $L$ has one of its ground state(s) with $\langle U_F\rangle\neq1+\mathscr{O}(1/L)$,
then it must be gapless.
Our strategy is to focus on the degree of spontaneous symmetry-breaking (SSB) of a lattice translation symmetry~\cite{Gioia:2022aa} and the approximate sign flipping property of $F$.
Since $F$ is winding-free,
it can be smoothly deformed to zero which enables us to derive infinitely many gaplessness indicators by static structure factors of arbitrary orders.
These structure-factor indicators potentially provide experimental approaches to detect gaplessness,
which cannot be achieved by earlier indicators~\cite{Su:2025aa,Tada:2026aa,Cheng:2026kch}.
Finally,
we apply analytic and numerical methods to test the efficiency of our results on gapless models and consistency in gapped models.

\paragraph{Preparations and the main results.---}
We consider a general Hamiltonian possessing U(1) symmetry defined on a one-dimensional lattice with a finite-dimensional local Hilbert space $\mathcal{H}_j$,
e.g., spin chains or fermion chains,
where $j$ labels the coordinate of the $j$-th unit cell.
U(1) symmetry is generated by $\bm{n}=\sum_{j}\hat{n}_j$
where the integer-valued $\hat{n}_j$ is the U(1)-charge operator of the unit cell $j$.
Here our Hamiltonians always possess some lattice translation $T$-symmetry relating unit cells $
T^{-1}\hat{n}_jT=\hat{n}_{j+1}$.
The existence of $T$ is necessary to define the thermodynamic limit by repeating the unit cells infinitely.
It should be noted that the unit cell or $T$-symmetry $\mathbb{Z}$ is highly model-dependent and we only make use of the existence of $T$,
\textit{without} specifying any microscopic data per unit cell.


%
%

A gapped Hamiltonian may have ground states with $T$-SSB in the following general form to one of $\mathbb{Z}$'s subgroup: 
\begin{eqnarray}
\mathbb{Z}\mapsto n_B\mathbb{Z},
\end{eqnarray}
where $n_B$ is some positive integer and $n_B=1$ trivially corresponds to the absence of $T$-SSB.

Our task is to recognize the gaplessness of the Hamiltonian with $L$ unit cells under a periodic boundary condition by the following general \textit{topologically trivial} twisting operator:
\begin{eqnarray}
{U}_F=\exp\left(i\sum_{m=1}^LF(m)\hat{n}_m\right),
\end{eqnarray}
where $F:\mathbb{Z}\rightarrow\mathbb{R}$ and satisfies
\begin{eqnarray}\label{trivial_twist}
F(m+L)=F(m),\,\, |F(m+1)-F(m)|=\mathscr{O}(1/L).
\end{eqnarray}
The typical topologically nontrivial twisting operator made use of by LSM theorem $F(m)=2\pi m/L$ does \textit{not} fulfill the above periodicity condition.

The periodicity of $F_m$ enables a discrete Fourier transformation:
\begin{eqnarray}\label{Fourier_F}
F(m)=\bar{F}+\sum_{p>0}\underbrace{a(p)\cos\left(\frac{2\pi pm}{L}\right)+b(p)\sin\left(\frac{2\pi pm}{L}\right)}_{\equiv F_p(m)},
\end{eqnarray}
and the derivative condition $|F(m+1)-F(m)|=\mathscr{O}(1/L)$ means that the Fourier coefficients $[a(p),b(p)]$ must converge to zero sufficiently fast as $p$ increases.
A Fourier-truncated $F$ ---
the $p$-component $F_p\neq0$ only when $|p|<p_\text{max}$ with a $L$-independent $p_\text{max}$ as $L\rightarrow\infty$ ---
always satisfies the above requirements.
Besides its obvious simplification,
it naturally has a well-defined thermodynamic limit for $F_m$ by
fixing $[a(p),b(p)]_{p=1,\cdots, p_\text{max}}$ as constants as $L\rightarrow\infty$.
We will use Fourier-truncated $F$ in most analyses below,
but we expect that
after careful treatment
our result can be also valid for general $F$ fulfilling Eq.~(\ref{trivial_twist})
as confirmed by our numerical study.

It has been shown that gapped ground states are almost $U_F$-invariant~\cite{Tasaki:2020aa}:

{\bf Lemma~1}: 
For a gapped U(1)-symmetric Hamiltonian $H$,
any of its normalized ground state $|\text{gs}\rangle$ after twisted by ${U}_F$ is ``almost'' a ground state:
\begin{eqnarray}
{U}_F|\text{gs}\rangle=|\Phi_\text{gs}\rangle+|\gamma_\text{ex}\rangle,
\end{eqnarray}
for some $|\Phi_\text{gs}\rangle$ in the ground-state sector
and $|\gamma_\text{ex}\rangle$ a state in the excited sector:
\begin{eqnarray}
&&\langle\Phi_\text{gs}|\Phi_\text{gs}\rangle=1+\mathscr{O}(1/L),\,\,\langle\gamma_\text{ex}|\gamma_\text{ex}\rangle=\mathscr{O}(1/L).\,\,\,\hfill\square
\end{eqnarray}
Actually,
the operator ${U}_F$ also almost preserves the norm of the excited spectrum:

{\bf Lemma~2: }For the gapped U(1)-symmetric Hamiltonian,
then,
any of its normalized excited state $|\text{ex}\rangle$ satisfies,
\begin{eqnarray}\label{op_ex}
{U}_F|\text{ex}\rangle=|\phi_\text{gs}\rangle+|\Gamma_\text{ex}\rangle,
\end{eqnarray}
where $|\phi_\text{gs}\rangle$ is in the ground-state sector and $|\Gamma_\text{ex}\rangle$ is in the excited spectrum satisfying:
\begin{eqnarray}
&&\langle\phi_\text{gs}|\phi_\text{gs}\rangle=\mathscr{O}(1/L),\,\,\,\langle\Gamma_\text{ex}|\Gamma_\text{ex}\rangle=1+\mathscr{O}(1/L).
\end{eqnarray}

\textit{Proof: }Since $U_F^\dagger=U_{-F}$ also satisfies {\bf Lemma~1},
we have $U_{-F}|\text{gs}\rangle=|\Phi_\text{gs}\rangle+|\gamma\rangle$ by the same notation in {\bf Lemma~1} for any normalized $|\text{gs}\rangle$.
Then any matrix element
$\langle\text{gs}|U_F|\text{ex}\rangle=\langle\gamma_\text{ex}|\text{ex}\rangle=\mathscr{O}(1/\sqrt{L})$ which means that the ground-state component $|\phi_\text{gs}\rangle$ of $U_F|\text{ex}\rangle$ satisfies $\langle\phi_\text{gs}|\phi_\text{gs}\rangle=\mathscr{O}(1/L)$.
Hence,
$\langle\Gamma_\text{ex}|\Gamma_\text{ex}\rangle=1+\mathscr{O}(1/L)$ since $|\text{ex}\rangle$ is normalized and $U_F$ is norm preserving.
\hfill$\square$\par

\textit{Illustrating theorem and main results.---}
To illustrate our basic idea,
we first give a Theorem which is mathematically clear although with redundant conditions.

{\bf Theorem 3: }
If a one-dimensional U(1)-symmetric gapped Hamiltonian with even length $L=2N$ has its lattice translation symmetry $\mathbb{Z}$ spontaneously broken to $n_B\mathbb{Z}\subseteq\mathbb{Z}$ and $n_B|N$ as $N\rightarrow\infty$,
any of the ground state(s) $|\text{gs}\rangle$ must satisfy
\begin{eqnarray}
\langle\text{gs}|U_F|\text{gs}\rangle=1+\mathscr{O}(1/L),
\end{eqnarray}
where $F$ \textit{additionally} satisfies
\begin{eqnarray}\label{add_F}
F(m)=-F({m+N}).
\end{eqnarray}
\textit{Proof: }
Defining $f\equiv F/2$ and
using {\bf Lemma~1} for $U_{f}$ and $U_{-f}=U^\dagger_f$,
we obtain that
\begin{eqnarray}
\left\{\begin{array}{l}U_f|\text{gs}\rangle=|\Psi_\text{gs}\rangle+|a_\text{ex}\rangle,\\
U^\dagger_f|\text{gs}\rangle=|\Gamma_\text{gs}\rangle+|b_\text{ex}\rangle,\end{array}\right.
\end{eqnarray}
with $\langle\Psi_\text{gs}|\Psi_\text{gs}\rangle\sim\langle\Gamma_\text{gs}|\Gamma_\text{gs}\rangle\sim1+\mathscr{O}(1/L)$ and $\langle a_\text{ex}|a_\text{ex}\rangle\sim\langle b_\text{ex}|b_\text{ex}\rangle\sim\mathscr{O}(1/L)$.

We define a large translation operator $\tau\equiv T^{N}$.
Due to the $T$-SSB nature,
any of the ground state(s) in the $T$-eigenspace, has $T$-eigenvalue in $\exp(i2\pi\mathbb{Z}/n_B)$,
so its $\tau$-eigenvalue is $1$ since $n_B|N$.
Therefore,
the arbitrary ground state $|\text{gs}\rangle$ satisfies
\begin{eqnarray}\label{tau_1}
\tau|\text{gs}\rangle=|\text{gs}\rangle.
\end{eqnarray}
The additional $F_m=-F_{m+N}$ gives
\begin{eqnarray}\label{uf_1}
U_f^{-1}=U_{-f}=\tau U_f\tau^{-1}.
\end{eqnarray}
Combining Eqs.~(\ref{tau_1},\ref{uf_1}),
we have
\begin{eqnarray}
1&=&\langle\text{gs}|\text{gs}\rangle\nonumber\\
&=&\langle\text{gs}|U_f\tau U_f\tau^{-1}|\text{gs}\rangle\nonumber\\
&=&\langle\text{gs}|U_f\tau (|\Psi_\text{GS}\rangle+|a_\text{ex}\rangle)\nonumber\\
&=&\langle\text{gs}|U_f|\Psi_\text{GS}\rangle+\langle\text{gs}|U_f\tau|a_\text{ex}\rangle.
\end{eqnarray}
On the other hand,
\begin{eqnarray}
\langle \text{gs}|U_fU_f|\text{gs}\rangle&=&\langle\text{gs}|U_f(|\Psi_\text{GS}\rangle+|a_\text{ex}\rangle)\nonumber\\
&=&\langle\text{gs}|U_f|\Psi_\text{GS}\rangle+\langle\text{gs}|U_f|a_\text{ex}\rangle\nonumber\\
&=&1-\langle\text{gs}|U_f\tau|a_\text{ex}\rangle+\langle\text{gs}|U_f|a_\text{ex}\rangle\nonumber\\
&=&1-\langle\text{gs}|\tau U_f^\dagger|a_\text{ex}\rangle+\langle\text{gs}|U_f|a_\text{ex}\rangle\nonumber\\
&=&1-\langle\text{gs}|U_f^\dagger|a_\text{ex}\rangle+\langle\text{gs}|U_f|a_\text{ex}\rangle\nonumber\\
&=&1-\langle a_\text{ex}|a_\text{ex}\rangle+\langle b_\text{ex}|a_\text{ex}\rangle\nonumber\\
&=&1+\mathscr{O}(1/L),
\end{eqnarray}
where we use the fact that $|\langle b_\text{ex}|a_\text{ex}\rangle|\leq\sqrt{\langle b_\text{ex}|b_\text{ex}\rangle\langle a_\text{ex}|a_\text{ex}\rangle}$.
It completes the proof of {\bf Theorem~3}.
\hfill$\square$\par

%
%
The requirements ``$L=2N$'',
``$n_B|N$'', and
``$F_m=-F_{m+N}$'' are physically unreasonable and make the practical use of the {\bf Theorem~3} nearly impossible,
especially ``$n_B|N$''.
Fortunately,
these conditions can be removed as in our main theorem below.

{\bf Theorem~4: }\textit{(Gaplessness indicator)}
For a one-dimensional U(1)-symmetric Hamiltonian,
any of the ground state(s) $|\text{gs}\rangle$ must satisfy
\begin{eqnarray}\label{thm_4}
\langle\text{gs}|U_F|\text{gs}\rangle=1+\mathscr{O}(1/L),
\end{eqnarray}
where $F_m$ satisfies that $F_{m+L}=F_m$ with a zero average value $\bar{F}=\sum_{m=1}^{L}F_m/L=0$.
If $|\text{gs}\rangle$ is a U(1)-eigenstate with U(1)-charge $Q$ and $\bar{F}\neq0$,
then
$\langle\text{gs}|U_F|\text{gs}\rangle=\exp(i\bar{F}Q)+\mathscr{O}(1/L)$.

This theorem gives us a gaplessness indicator;
once any ground state does not satisfy Eq.~\eqref{thm_4},
its parent Hamiltonian must be gapless.
Since the case with $\bar{F}\neq0$ is straightforward to deal with,
let us assume $\bar{F}=0$.
Before we prove {\bf Theorem~4},
we make several useful observations to sketch the proof. 
By Eq.~\eqref{Fourier_F}
we conclude that each $p$-component \textit{almost} flips its sign through:
\begin{eqnarray}
F_p[{m+n(p)}]=-F_p({m})+\mathscr{O}(1/L),
\end{eqnarray}
where
\begin{eqnarray}
N_p\equiv \left\lfloor\frac{L}{2p}\right\rfloor,
\end{eqnarray}
which approximately plays the role of $N$ before in Eq.~\eqref{add_F}.
In addition,
we know $N_p\in n_B\mathbb{Z}+l_p$ with a unique integer $l_p\in[0,n_B)$ which is bounded by the finite number $n_B$.
Since $L\rightarrow\infty$,
this bounded remainder $l_p< n_B\ll L$ is under control,
which plays the role of ``$n_B|N$'' before.

We do a Fourier transformation of the local U(1) charge:
\begin{eqnarray}\label{def_fourier}
\hat{n}_m=\frac{1}{\sqrt{L}}\sum_pn(p)\exp(i2\pi pm/L).
\end{eqnarray}

Since we will make the above observations work by several technical Lemmas,
first-time readers may proceed directly to {\bf Theorem~9}.

{\bf Lemma~5: }\textit{(Vanishing $\langle n({p\neq0})\rangle$)}
For a gapped U(1)-symmetric Hamiltonian,
then $n({p})$ with a \textit{fixed} $p\neq0$ is strictly zero in the ground state sector when $L$ is sufficiently large.

\textit{Proof: }Let us choose $T$-eigenstate basis for the ground-state sector and consider
the matrix element of $n({p\neq0})$ between any pair of ground states with lattice momenta,
say $\exp(i2\pi p_1/n_B)$ and $\exp(i2\pi p_2/n_B)$ with $p_{1,2}\in\mathbb{Z}$~\cite{Gioia:2022aa}.
The operator $n(p):T^{-1}n(p)T=n(p)\exp(i2\pi p/L)$ carries a lattice momentum $\exp(i2\pi p/L)$.
Thus, the difference $\exp[i2\pi (p_1-p_2)/n_B]$ cannot be cancelled by $\exp(i2\pi p/L)$ when $L> (p_1-p_2)pn_B$.
Thus, $n_{p\neq0}|_\text{gs sector}=0$ as $L\rightarrow\infty$.
\hfill$\square$\par

We extract out the expected property of $U_F$ in {\bf Theorem~4} by a useful definition:

{\bf Definition: }\textit{(Almost-identity operator)} A unitary operator $W$ is an almost-identity operator if $\langle\text{gs}|W|\text{gs}\rangle=1+\mathscr{O}(1/L)$.

By the norm preservation and unitarity of $W$,
we have
\begin{eqnarray}\label{Wdagger}
\left\{\begin{array}{l}W|\text{gs}\rangle=[1+\mathscr{O}(1/L)]|\text{gs}\rangle+|\Phi\rangle,\\
W^\dagger|\text{gs}\rangle=[1+\mathscr{O}(1/L)]|\text{gs}\rangle+|\Phi'\rangle,\end{array}\right.
\end{eqnarray}
where $|\text{gs}\rangle$ is any normalized ground state and $\langle\Phi|\Phi\rangle=\mathscr{O}(1/L)$ with $\langle\Phi|\text{gs}\rangle=0$,
and similarly for $|\Phi'\rangle$.
Here $|\Phi\rangle$ and $|\Phi'\rangle$ can be a superposition of ground states and excited states.
Almost-identity operator only appears as an identity in the ground-state sector.
In the excited-state sector,
it almost preserves the norm shown by Eq.~(\ref{Wdagger}) generalizing {\bf Lemma~2}:

{\bf Lemma~6: }\textit{(Multiplicativity)}
For an almost-identity operator $W$ and any normalized excited state $|\text{ex}\rangle$,
we have $W|\text{ex}\rangle=|\Gamma_\text{ex}\rangle+|\phi_\text{gs}\rangle$ with $|\Gamma_\text{ex}\rangle:\langle\Gamma_\text{ex}|\Gamma_\text{ex}\rangle=1+\mathscr{O}(1/L)$ an excited state and $|\phi_\text{gs}\rangle:\langle\phi_\text{gs}|\phi_\text{gs}\rangle=\mathscr{O}(1/L)$ a ground state.
A multiplication of almost-identity operators $(W_1W_2)$ is still almost-identity by considering $\langle\text{gs}|W_1W_2|\text{gs}\rangle$ and $W_1^\dagger|\text{gs}\rangle$ in Eq.~\eqref{Wdagger}.
\hfill$\square$\par

The following Lemma is essential for the final proof.

{\bf Lemma~7: }For a gapped U(1)-symmetric Hamiltonian and a periodic function $s$ satisfies $s(m)=O(1/L)$ and $s({p=0})=0$,
the operator $U_{s}$ is almost-identity.

\textit{Proof: }
We choose $\{|\text{gs}_j\rangle\}$ that satisfying the clustering principle, as the basis of the ground state sector.
These ground states are connected by $T$ transformation and are superpositions of distinct $T$-eigenstates in the case of $T$-SSB ($n_B>1$).
Similarly,
we take a Fourier truncated function $s$: $s(p)=0$ when $|p|>p_\text{max}$.

Let us consider $\langle\text{gs}_j|U_s|\text{gs}_j\rangle$.
Since $s({p=0})=0$,
we obtain $\langle\text{gs}_j|U_s|\text{gs}_j\rangle=\langle\exp[i\sum_ms(m)\delta \hat{n}_m]\rangle$ with $\delta \hat{n}_m\equiv \hat{n}_m-\langle\text{gs}_j|\hat{n}_m|\text{gs}_j\rangle$ and this subtraction does not affect its nonzero momenta: $\delta n({p\neq0})=n(p)$.

Then we do a Taylor expansion:
\begin{eqnarray}
&&\langle\text{gs}_j|U_s|\text{gs}_j\rangle=1+i\sum_ms(m)\left\langle \delta \hat{n}_m\right\rangle\nonumber\\
&&+\sum_{k>1}\frac{i^k}{k!}\sum_{m_{1,\cdots,k}}s(m_1)\cdots s(m_k)\left\langle\delta \hat{n}_{m_1}\cdots\delta \hat{n}_{m_k}\right\rangle\nonumber\\
&=&1+\sum_{k>1}\frac{i^k}{k!}\sum_{m_{1,\cdots,k}}s(m_1)\cdots s(m_k)\langle\delta n_{m_1}\cdots\delta n_{m_k}\rangle,\nonumber
\end{eqnarray}
since $\sum_ms(m)\left\langle \delta \hat{n}_m\right\rangle=\sqrt{2L}\sum_{p\neq 0}s({-p})\langle n(p)\rangle=0$ by {\bf Lemma~5}.
Because the system has a nonzero spectral gap and $\langle\delta \hat{n}_m\rangle=0$,
the ground-state \textit{fluctuation} correlation $\langle\delta \hat{n}_{m1}\cdots \delta \hat{n}_{m_k}\rangle$ is exponentially suppressed by distances unless each operator therein has at least one another operator within the correlation length.
Such a consideration significantly suppresses the original summation $\sum_{m_1,\cdots,m_k}\sim L^k$ effectively down to $\frac{L^{\lceil k/2\rceil}k!}{2^{\lceil k/2\rceil}\lceil k/2\rceil!}$ with the leading order contributed by the pairing clusterings $\delta\hat{n}_{m_i}\delta\hat{n}_{m_j}$.
Therefore,
by $s(m)=\mathscr{O}(1/L)$ and the operator $\delta \hat{n}_m$ is bounded, say by $B$, due to the finite-dimensional local Hilbert space,
\begin{eqnarray}
\langle\text{gs}_j|U_s|\text{gs}_j\rangle&\simeq&1+\sum_{k>1}\frac{B^k}{2^{\lceil k/2\rceil}\lceil k/2\rceil!}\mathscr{O}\left(\frac{1}{L^{\lfloor k/2\rfloor}}\right)\nonumber\\
&=&1+\mathscr{O}\left(\frac{1}{L}\right).
\end{eqnarray}
When $k= \mathscr{O}(L)$,
the prefactor $\frac{B^k}{2^{\lceil k/2\rceil}\lceil k/2\rceil!}\sim\frac{1}{L^{\mathscr{O}(L)}}$ is sufficiently suppressed and does not contribute to the estimate above.
It completes the proof of {\bf Lemma~7}.
\hfill$\square$\par

Since all the ground states have lattice momentum valued in $\exp(i2\pi \mathbb{Z}/n_B)$~\cite{Gioia:2022aa} and $N_p\in n_B\mathbb{Z}+l_p$,
we obtain
\begin{eqnarray}\label{gs_tau}
\tau_p|_\text{gs sector}=T^{l_p},
\end{eqnarray}
where $\tau_p\equiv T^{N_p}$ in an analog of $\tau$ before in {\bf Theorem~1}.

We focus on a single fixed $p$-component:

{\bf Lemma~8: }For a gapped U(1)-symmetric Hamiltonian and $f_p\equiv F_p/2$ [which are functions as in Eq.~\eqref{Fourier_F} rather than Fourier coefficients], 
\begin{eqnarray}
W_{f_p}\equiv U_{f_p}^{-1}(\tau_p T^{-l_p})U_{f_p}
\end{eqnarray}
is almost-identity.

\textit{Proof: }
We choose any orthogonal basis of the ground state sector denoted as $\{|\text{gs}_j\rangle\}$.
Combining {\bf Lemma~1}, Eq.~(\ref{gs_tau}),
and the fact that $\tau_p T^{-l_p}$ as a symmetry of the Hamiltonian cannot map any excited state to the ground state sector,
we obtain that the diagonal matrix elements $\langle \text{gs}_j|W_{f_p}|\text{gs}_j\rangle=1+\mathscr{O}(1/L)$.
Using the notation in the {\bf Lemma~1},
we have the off-diagonal element $\langle \text{gs}_{i\neq j}|W_{f_p}|\text{gs}_j\rangle=\langle \Phi_{\text{gs}_i}|\Phi_{\text{gs}_j}\rangle+\mathscr{O}(1/L)=-\langle\gamma_i|\gamma_j\rangle+\mathscr{O}(1/L)=\mathscr{O}(1/L)$,
which completes the proof of {\bf Lemma~8}.
\hfill$\square$\par

\paragraph{Proof of {\bf Theorem~4}.---}
We define $g_p$ through $U_{g_p}\equiv\tau_p U_{f_p}\tau_p^{-1}$.
Then $\tilde{s}_p\equiv g_p+f_p$ satisfies the condition of {\bf Lemma~7} since $g_p$ is \textit{almost} $-f_p$.
Making a $l_p$-translation $\tilde{f}_p:\tilde{f}_p(m)\equiv f_p(m+l_p)$,
we obtain that for any ground state $|\text{gs}\rangle$
\begin{eqnarray}
U_{\tilde{s}_p}|\text{gs}\rangle&=&U_{F_p}U^{-1}_{f_p}\tau U_{f_p}\tau^{-1}|\text{gs}\rangle\nonumber\\
&=&U_{F_p}U_{f_p}^{-1}\tau T^{-l_p}T^{l}U_{f_p} T^{-l_p}|\text{gs}\rangle\nonumber\\
&=&U_{F_p}U_{f_p}^{-1}\tau T^{-l}U_{\tilde{f}_p}|\text{gs}\rangle\nonumber\\
&=&U_{F_p}\left(U_{-f_p}U_{\tilde{f}_p}\right)\left[U_{\tilde{f}_p}^{-1}\left(\tau_p T^{-l_p}\right)U_{\tilde{f}_p}\right]|\text{gs}\rangle\nonumber\\
&=&U_{F_p}U_{s_p}W_{\tilde{f}_p}|\text{gs}\rangle,
\end{eqnarray}
where $s_p\equiv \tilde{f}_p-f_p$ satisfies the condition of {\bf Lemma~7} since $l_p$ is bounded by the finite number $n_B$.
Grouping the $p$-components gives and restricting to $F_{p=0}=\bar{F}=0$,
\begin{eqnarray}
\prod_{p>0}U_{\tilde{s}_p}|\text{gs}\rangle=\prod_{p>0}U_{F_p}U_{s_p}W_{\tilde{f}_p}|\text{gs}\rangle=U_{F}\prod_{p>0}U_{s_p}W_{\tilde{f}_p}|\text{gs}\rangle,\nonumber
\end{eqnarray}
which completes the proof of {\bf Theorem~4} by applying {\bf Lemma~6} to the above multiplication,
{\bf Lemma~7} to $U_{s_p}$ and {\bf Lemma~8} to $W_{\tilde{f}_p}$.
\hfill$\square$\par

\paragraph{Physical consequences and predictions.---}
The topologically trivial $F$ enables us to have $\langle U_{tF}\rangle$ to obey {\bf Theorem~4} for any small $t$,
which enables us to do a $t$-expansion after cancelling $t^0$:
\begin{eqnarray}
\mathscr{O}\left(\frac{1}{L}\right)=\sum_{m=1}^\infty \sum_{\{k_i\neq0\}}t^M\left(\cdots\right)_{\{k\}}\sqrt{L^M}\langle n_{k_1}\cdots n_{k_M}\rangle,\nonumber
\end{eqnarray}
where $\left(\cdots\right)_{\{k\}}\sim\mathscr{O}(1)$ is various multiplications of $a(k_i)$ and $b(k_j)$.
Since $t$ is arbitrarily chosen,
we have:

{\bf Theorem~9 } \textit{(Static structure factor indicators)} 
For a U(1)-symmetric gapped Hamiltonian and any integer $M$,
the $M$-th order static structure factor satisfies for any of its ground state(s) $|\text{gs}\rangle$:
\begin{eqnarray}\label{thm_9}
\langle \text{gs}|n({k_1})\cdots n({k_M})|\text{gs}\rangle=\mathscr{O}\left(\frac{1}{L^{1+\frac{M}{2}}}\right),
\end{eqnarray}
for any fixed nonzero $\{k_j\}$ as $L\rightarrow\infty$.\hfill$\square$\par

To relate it to the real experiment,
we notice that the continuous momenta corresponding to $k_j$ is $K_j={2\pi k_j}/{L}$ by Eq.~\eqref{def_fourier}.
Thus,
{\bf Theorem~9} should be re-interpreted experimentally as
\begin{eqnarray}\label{thm_9_2}
\langle n(K_1)\cdots n(K_M)\rangle=\mathscr{O}\left(K^{1+\frac{M}{2}}\right),\,\,\text{as }K\rightarrow0^+,
\end{eqnarray}
if nonzero $|K_j|\sim K$.
This statement restricts the form of nonlinear responses detected by,
e.g., neutron scattering experiments at non-zero scattering momenta~\cite{Zaliznyak:1994aa,Lee:2002aa,Zaliznyak:2004aa,Muhlbauer:2019aa}.

%

\begin{figure}[h] 
\centering 
\includegraphics[width=0.5\textwidth]{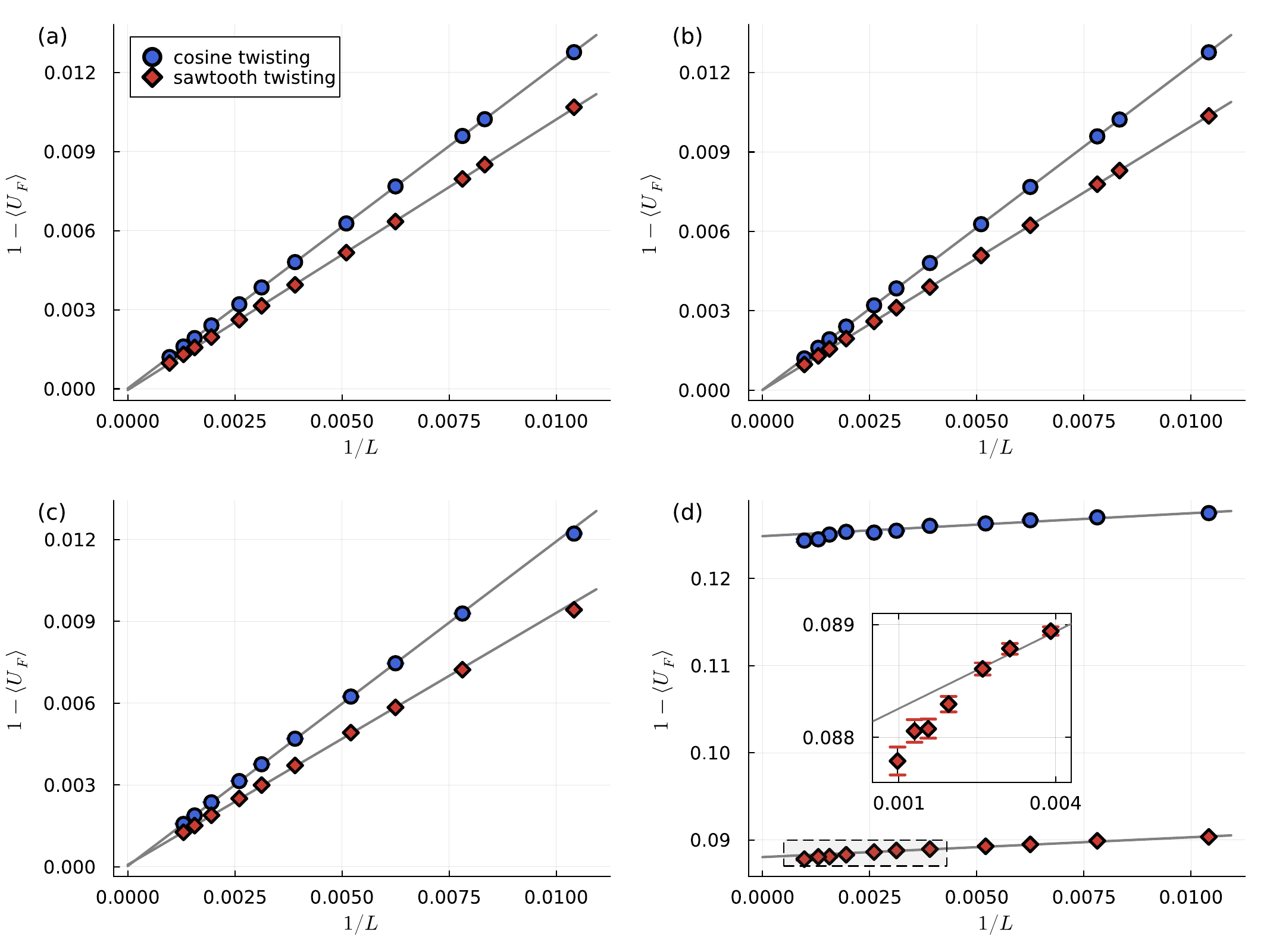} 
\caption{QMC simulations of $(1-\langle U_\text{F}\rangle)$ with $F_\text{cos}$ and $F_\text{sawtooth}$: (a) the gapped AKLT model; (b) the MG model; (c)  XXZ model with $\Delta=2$ in the gapped regime; (d) the gapless spin-1/2 HAF model.}\label{numerics}
\end{figure}

Let us apply our {\bf Theorems~4,9} to concrete systems:

\begin{itemize}
\item Fermi sea:

For a general one-dimensional spinless Fermi sea:
\begin{eqnarray}
|\text{gs}\rangle=\prod_{n=1}^{n_\text{D}}\prod_{k:k_\text{n,L}\leq k\leq k_\text{n,R}}c^\dagger_k|\text{vac}\rangle,
\end{eqnarray}
we calculate that $(\bar{F}=0)$
\begin{eqnarray}
&&\langle U_{tF}\rangle=\left\langle \exp\left[\frac{t}{2}\sum_{p>0}[a(p)+ib(p)]c^\dagger_kc_{k-\frac{2\pi}{L}p}-\text{h.c.}\right]\right\rangle\nonumber\\
&=&1-\frac{t^2}{4}n_\text{D}\sum_{p=1}^{p_\text{max}}p[a(p)^2+b(p)^2]+O(t^4)\neq1+\mathscr{O}(1/L),\nonumber
\end{eqnarray}
which means that the Fermi sea cannot be a ground state of any gapped fermion chain.

\item {Two-point correlators in Luttinger liquid}

The Luttinger liquid described by a bosonic field $\varphi$ is gapless since
the Fourier transformation of its local U(1) charge
$\partial_x\varphi$ satisfies
$\langle\partial\varphi(q)\partial\varphi(-q)\rangle\sim |q|$ as $q\rightarrow0^+$.
On the other hand of a massive free boson with mass $m$,
$\langle\partial\varphi(q)\partial\varphi(-q)\rangle\sim\frac{q^2}{\sqrt{q^2+m^2}}\sim q^2$ obeys Eq.~\eqref{thm_9_2}.
Nevertheless,
the field-theoretical local Hilbert space is infinitely dimensional,
so whether our theorems can be rigorously applicable here is unclear.

\item{Spin models}

Let us use two following $F$ functions in $m\in[1,L]$:
\begin{eqnarray}
&&F_\text{cos}(m)=\cos\frac{2\pi m}{L};\,\,F_\text{sawtooth}(m)=1-\left|\frac{4m}{L}-2\right|,\nonumber
\end{eqnarray}
to (spin-1)
AKLT model $H_\text{AKLT}=\sum_i\vec{S}_i\cdot\vec{S}_{i+1}+\left(\vec{S}_i\cdot\vec{S}_{i+1}\right)^2/3$, (spin-1/2) Majumdar-Ghosh (MG) model $H_\text{MG}=\sum_i\vec{S}_i\cdot\vec{S}_{i+1}+
\frac{1}{2}\vec{S}_i\cdot\vec{S}_{i+2}$, (spin-1/2) XXZ model $H_\text{XXZ}=\sum_iS^x_{i}S^x_{i+1}+S^y_{i}S^y_{i+1}+\Delta S^z_{i}S^z_{i+1}$, and (spin-1/2) Heisenberg antiferromagnetic (HAF) model $H_\text{HAF}=\sum_i\vec{S}_i\cdot\vec{S}_{i+1}$.
We obtain the numerical results FIG.~\ref{numerics} by QMC~\cite{Sandvik:1991aa,Sandvik:1992aa,Syljuasen:2002aa},
in which all those gapped Hamiltonians exhibit $\langle U_F\rangle=1+\mathscr{O}(1/L)$ and the gapless Hamiltonians are well diagnosed by our method.
The $F_\text{sawtooth}$ is not Fourier-truncated but satisfies Eq.~\eqref{trivial_twist},
so the consistent numerical result indicates our {\bf Theorems} are probably applicable for such general $F$'s.
\end{itemize}

\paragraph{Conclusions and Discussions.---}In this work,
we make use of topologically trivial twisting operators to indicate gaplessness of U(1)-symmetric systems without requiring stronger symmetry.
It further enables us to propose experimental access to diagnose gaplessness through static structure factors.
These two aspects cannot be achieved by the previous topologically nontrivial twisting operators.

Moreover,
we claim below,
even though in the presence of stronger symmetry,
our topologically trivial indicators should be more efficient than those nontrivial ones.
The important aspect of nontrivial winding is its potentially nontrivial lattice momentum,
which is crucial in the LSM theorem.
However,
in the gaplessness indicator,
the winding of the twisting operator is chosen so that its momentum is \textit{trivial}.
Therefore,
such topological nontriviality is not substantial.
On the other hand,
our indicator has no definite lattice momentum,
which is actually a \textit{good} property to detect gaplessness;
$U_F|\text{gs}\rangle$ will involve many low-lying energy states with various momenta to suppress $\langle\text{gs}|U_{F}|\text{gs}\rangle$,
which explains its efficiency fundamentally.

\paragraph{Acknowledgements.---}The authors thank Akira Furusaki for helpful discussions.
	The work of Y.~Y. was supported by the National Key Research and Development Program of China (Grant No.~2024YFA1408303),
	the National Natural Science Foundation of China (Grant No.~12474157),
	the sponsorship from Yangyang Development Fund,
	and Xiaomi Young Scholars Program. L.H.L. acknowledges the partial support from a Quantum SuperSEED grant (ICDS\_QS25\_029093) from the Institute for Computational and Data Sciences at the Pennsylvania State University and support from a startup
fund from the Pennsylvania State University (Zhen Bi). F.-F.~S. is supported by Grant-in-Aid for JSPS Fellows (Grant No.~25KF0183).
	Part of the computation in this work has been done using the facilities of the Supercomputer Center, the Institute for Solid State Physics, the University of Tokyo.

\bibliographystyle{apsrev4-1}

%

---------------------------

\end{document}